\documentclass[]{aa}
\usepackage{amsmath} 
\usepackage{amssymb}

\usepackage[utf8]{inputenc}
\usepackage{epsfig}
\usepackage{url}
\usepackage{color}
\definecolor{darkgreen}{rgb}{0,.6,0}
\definecolor{linkcol}{rgb}{.6,0,0}
\usepackage[colorlinks=true,citecolor=darkgreen]{hyperref}

\usepackage{natbib}
\bibpunct{(}{)}{;}{a}{}{,}
\usepackage{subfigure}

\usepackage{upgreek}  
\newcommand\unit[1]{\,{\rm #1}} 

\def\eg{{\it e.g.}}
\def\ie{{\it i.e.}}

\def\lta{~\raise.4ex\hbox{$<$}\llap{\lower.6ex\hbox{$\sim$}}~}
\def\gta{~\raise.4ex\hbox{$>$}\llap{\lower.6ex\hbox{$\sim$}}~}

\def\msol{\ensuremath{\rm \mass_\odot}} 
\newcommand\solm\msol

\usepackage{dcolumn} \newcolumntype{d}[1]{D{.}{.}{#1}}
\usepackage{multirow}

\bibpunct{(}{)}{;}{a}{}{,}

\begin{document}

\title{Comets as collisional fragments of a primordial planetesimal disk}
\titlerunning{Comets as collisional fragments}

\author{A. Morbidelli\inst{1}\and H. Rickman\inst{2,3}}

\authorrunning{Morbidelli and Rickman}

\institute{
D\'epartement Lagrange, University of Nice -- Sophia Antipolis, CNRS, Observatoire de la C\^{o}te d'Azur, Nice, France
\and P.A.S. Space Research Center, Bartycka 18A, PL-00-716 Warszawa, Poland
\and Dept. of Physics and Astronomy, Uppsala University, Box 516, SE-75120 Uppsala, Sweden
 }

\abstract{The Rosetta mission and its exquisite measurements have revived the debate on whether comets are pristine planetesimals or collisionally evolved objects.} 
{We investigate  the collisional 
evolution experienced by the precursors of current comet nuclei during the early stages of the 
Solar System, in the context of the so-called ``Nice Model''.} 
{We consider two environments for the collisional evolution: (1) the trans-planetary planetesimal disk, from the time of gas removal until the disk was dispersed by the migration of the ice giants, and 
(2) the dispersing disk during the time that the scattered disk was formed. Simulations have been 
performed, using different methods in the two cases, to find the number of destructive collisions 
typically experienced by a comet nucleus of 2\unit{km} radius.} 
{In the widely accepted scenario, where the dispersal of the planetesimal disk occurred at the time 
of the Late Heavy Bombardment about 4\unit{Gy} ago, comet-sized planetesimals have a very small  
chance to survive against destructive collisions in the disk. On the extreme assumption that the 
disk was dispersed directly upon gas removal, there is a chance for a significant fraction of the 
planetesimals to remain intact. However, these survivors would still bear the marks of many 
non-destructive impacts.} 
{The Nice Model of Solar System evolution predicts that typical km-sized comet nuclei are 
predominantly fragments resulting from collisions experienced by larger parent bodies. An 
important goal for further research is to investigate, whether the observed properties of comet 
nuclei are compatible with such a collisional origin.} 


\date{[Received / accepted]}

\maketitle
\section{Introduction}\label{sec:intro}

Comet nuclei are usually thought to be icy planetesimals, formed beyond the snow-line in the nascent 
Solar System. As such, they are naturally considered as precious targets of space missions -- \eg, {\it 
Rosetta}. This concept is supported by the properties of comet nuclei derived from observations. The 
low bulk densities (\eg, Rickman, 1989, Davidsson et al., 2007), the negligible tensile strength 
inferred for comet D/Shoemaker-Levy 9 (Asphaug and Benz, 1996), {and the low tensile strength of the surface layer required to explain their activity (Blum et al., 2014)} are consistent with low-velocity 
accretion, in line with the general expectation for planetesimals formed at large distance from the Sun. Many 
comets have exhibited evidence for a very important contribution by the super-volatile CO molecule 
to their outgassing activity (\eg, Bockel\'ee-Morvan et al., 2004). This may be taken as an indication 
of a chemically pristine nature of the 
material that comet nuclei are made of, which supports the idea of a very gentle accretion process.

However, the issue of collisional evolution in the population of icy planetesimals has also been brought 
to light in several papers. Davis and Farinella (1997) modeled the collisional evolution of the 
Edgeworth-Kuiper Belt (EKB), which at the time was thought to be the source region for the Jupiter family 
comets (JFCs). They found that, while large EKB members are likely primordial objects, those with radii 
of a few km, like typical JFCs, are multi-generational fragments formed by the splitting of larger objects (see also Schlichting et al., 2013). 
Simultaneous with this first investigation, the Scattered Disk was discovered (Luu et al., 1997) and it was rapidly recognized to be a more efficient source of JFCs than the EKB. Due to the longer orbital periods of its typical orbits, the Scattered Disk is believed to be less collisionally evolved (Rickman, 2004), so that one could think that the observed JFCs have a higher chance to be primordial planetesimals than in the Farinella and Davis (1997) analysis.

Another scenario was considered by Stern and Weissman (2001): the formation of the Oort cloud. This 
was modeled in the classical picture of gravitational ejection of icy planetesimals from the growth 
region of the giant planets (Safronov, 1977). Stern and Weissman showed that, during the course of this process, 
comet nuclei would be destroyed by collisions with small debris. The authors concluded that the 
storage into the Oort cloud would be delayed until the comet source region had been cleared of 
material so that the collisional lifetime becomes longer than the ejection lifetime. Naturally, in this 
scenario most of the Oort cloud comets would still bear the marks of collisional erosion. A similar conclusion 
would apply also to the origin of the Scattered Disk. 

Charnoz and Morbidelli (2003, 2007) -- CM03/07 hereafter -- introduced a new algorithm for evaluating the effects of collisions in a 
population of small bodies subject to a complex and rapid dynamical evolution through gravitational 
perturbations, as is the case for planetesimals ejected from the giant planets region. This replaced the particle-in-a-box 
models earlier used. CM03/07 showed that, for some appropriate initial size distributions (similar to that currently observed in the trans-Neptunian region) a sufficiently large number of 
comet-size bodies would have reached the Oort cloud and the Scattered Disk. Although not explicitly discussed in these papers, they found that the vast majority of the Oort Cloud objects larger than 1\unit{km} in radius would be pristine planetesimals. However, in the Scattered Disk, only 2\% of the final population of objects of this size would be primordial, the rest being collisional fragments.

Since these papers were produced, the {\it Nice Model} for the early evolution of the Solar System has been 
introduced (Tsiganis et al., 2005; for the latest version, see Levison et al., 2011). This changes the 
picture of the origin and evolution of comets in important ways. One central concept is that of the trans-planetary 
disk of icy planetesimals. This was the disk of objects formed during the infant stages of the Solar System beyond the original orbits of all giant planets, which were originally closer to the Sun. This disk extended out to about 30\unit{AU} and had a total mass of 20--50 Earth masses. It remained relatively quiescent until it was eventually dispersed as a consequence of a dynamical instability among the giant planets and of the planets' subsequent migration towards their current orbits. The trans-planetary  disk, upon its dispersal, should have given rise to both the Scattered Disk and the Oort cloud (Brasser and  Morbidelli, 2013). Thus, this disk  may once have been the repository for all the comets observed today. This 
would be compatible with the lack of evidence for any clear-cut differences in molecular composition 
(A'Hearn et al., 2012) or D/H ratios (Altwegg et al., 2015) between JFCs and comets of Oort Cloud provenance, \ie, long-period comets (LPCs) and Halley-type comets (HTCs). Notice that there is today no alternative model capable of fully explaining the structure of the outer Solar System without invoking a Nice-model-like instability of the giant planets associated with the dispersal of the trans-planetary disk.

In this paper we investigate the collision rates involving the members of the trans-planetary disk 
during its whole evolution. First, we make the standard assumption that the dispersal of the disk coincided 
with the beginning of the so-called Late Heavy Bombardment (Gomes et al., 2005; Morbidelli et al., 2012), which means 
a lifetime of about {450}\unit{Myr} for the disk, prior to its dynamical dispersal. It is likely that the dynamical state of the disk was significantly excited. In fact, the probability that an object survived the dynamical {dispersal} of the disk, remaining permanently trapped into the hot EKB population (including mean motion resonances with Neptune) is less than $10^{-3}$ (see Nesvorn\'y, 2015 for the most updated estimate); this means that about $1\,000$ Pluto-size objects should have existed in the primordial disk (Stern, 1991). These bodies would have induced a significant excitation in the disk, causing a velocity dispersion of the order of $0.5-1$\unit{km/s} (Levison et al., 2011). If the disk stayed in this state for hundreds of millions of years, it is likely that the collisional evolution of comet-size objects has been severe. We will quantify this in Sect.~\ref{pre-instability}. The conclusions will apply to both comets in the Scattered Disk and in the Oort Cloud, given that the trans-planetary disk was the progenitor of both these reservoirs (Brasser and Morbidelli, 2013). 

However, it is not yet certain that the dynamical dispersal of the trans-planetary disk occurred late. The formation of the latest basins on the Moon (Imbrium and Orientale, and possibly all Nectarian basins) requires that new projectiles appeared in the terrestrial planet crossing region several 100\unit{My} after terrestrial planet formation (Bottke et al., 2007). The late instability of the giant planets would do this in a natural way (Bottke et al., 2012; Morbidelli et al., 2012). However, alternative models have been proposed. Some (\'Cuk, 2012; Minton et al., 2015) invoke unlikely spectacular collision events in the inner Solar System, generating a large amount of debris which would have subsequently bombarded the Moon and the terrestrial planets. These models have not yet been thoroughly tested against all available Solar System constraints. Other models, such as the destabilization of a population of lunar coorbital objects (\'Cuk and Gladman, 2006) or of a fifth terrestrial planet that would have then partially destabilized the asteroid belt (Chambers, 2007) did not pass such tests (\'Cuk and Gladman, 2009; Brasser and Morbidelli, 2011). The main argument independent of the lunar crater record in favor of a late dispersal of the trans-planetary disk is that the impact basins on Iapetus (a satellite of Saturn) have topographies that have relaxed by 25\% or less, which argues that the surface layer of Iapetus was already very viscous at the time of basin formation. According to models of the thermal evolution of the satellite, this high viscosity could not be possible earlier than 200\unit{My} after the beginning of the Solar System (Robuchon et al., 2011), which implies that the basins of Iapetus formed late. Nevertheless, this constraint remains model-dependent. 

Thus, in the second part of the paper we consider the case of an early dispersal of the trans-planetary disk, occurring just after gas removal. In this case, the collisional evolution prior to the instability would be negligible (the disk would have survived just a few My and, presumably, its planetesimals would have had a very small velocity dispersion due to gas drag); however, during the dispersal of the disk, the collisional evolution might have been severe (similar to the case studied by Stern and Weissman, 2001). In Sect.~\ref{dispersal} we quantify the collisional evolution of comet-size bodies eventually stored in the Scattered Disk during such dispersal. Obviously, if the trans-planetary disk dispersed late, the real collisional evolution of comets now in the Scattered Disk would be the sum of those studied in Sects.~\ref{pre-instability} and~\ref{dispersal}.   

This paper is structured as follows. We start presenting the logic of our reasoning, our methods, assumptions and choice of parameters in Sect.~\ref{methods}; the results are presented in the aforementioned Sects.~\ref{pre-instability} and~\ref{dispersal}; a discussion and summary of conclusions are given in Sect~\ref{conclusions}.

\section{Methods and principles}
\label{methods}

In this work we basically follow the approach of CM03/07, but with some important variants detailed in this Section. The CM03/07 approach is very suitable to study the collisional evolution of a population of bodies undergoing a significant dynamical evolution, with orbital histories that can be quite different from one particle to the other. It basically consists of three steps. First, one does a numerical simulation where the dynamical history of each particle is followed. Second, from the numerical simulation output, one computes the intrinsic collision probability and impact velocity of each particle with all others at each output time-step. Third, one assumes that each particle in the simulation is a tracer of a swarm of particles with a given initial size distribution. Then, using the information computed in the second step, one evolves the size distributions associated to each tracer from one output step to another. This involves computing the minimal projectile size for a catastrophic impact on targets of any given size, and the size distribution of the generated fragments. 

Below, we detail how we performed each of these three steps and in particular how we simplified the third step in order to reduce the parameter space we need to explore, although still satisfying our needs and achieving our goals. 

\begin{figure}[t!]
\centerline{\includegraphics[width=6.cm,angle=-90]{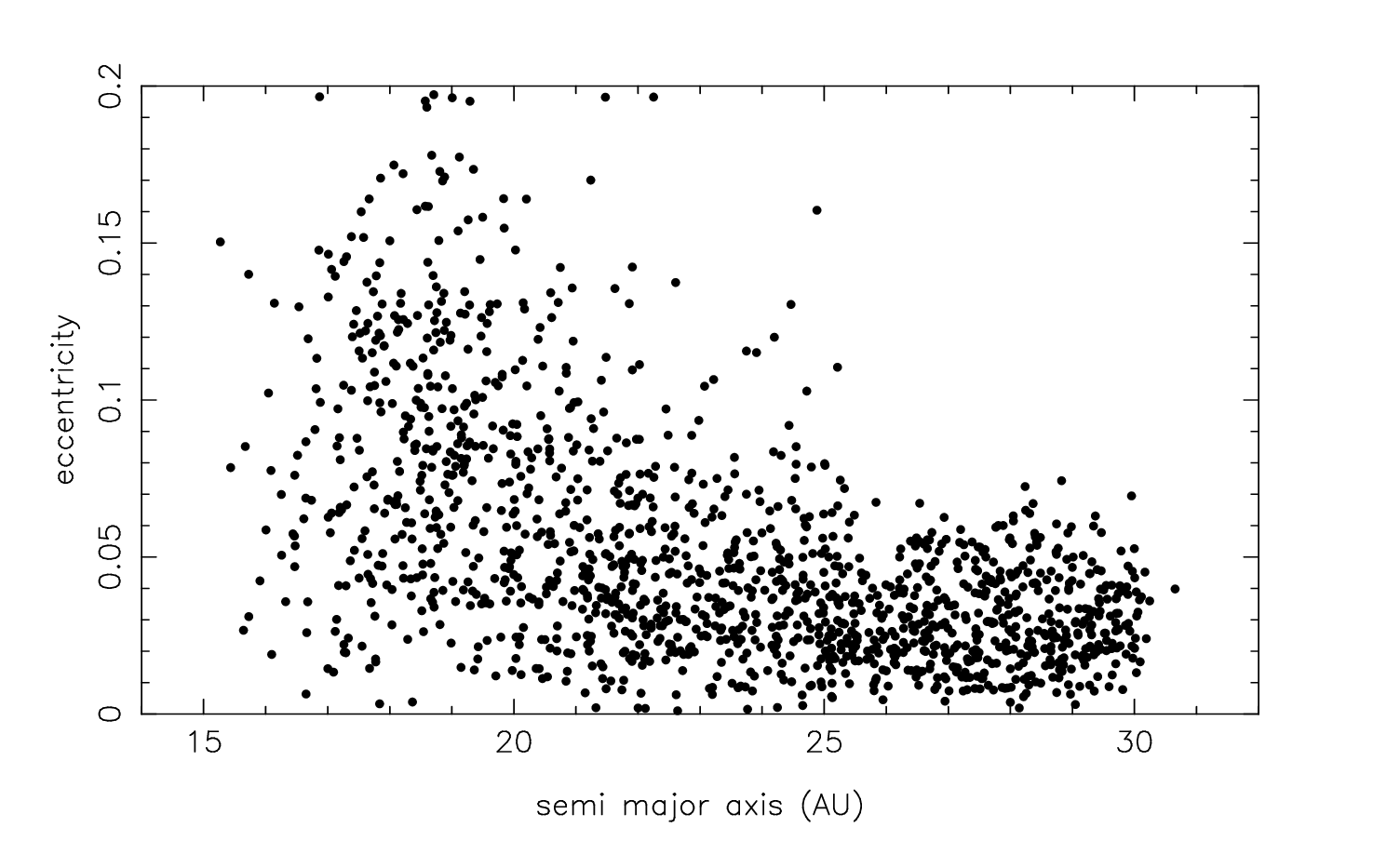}}
\centerline{\includegraphics[width=6.cm,angle=-90]{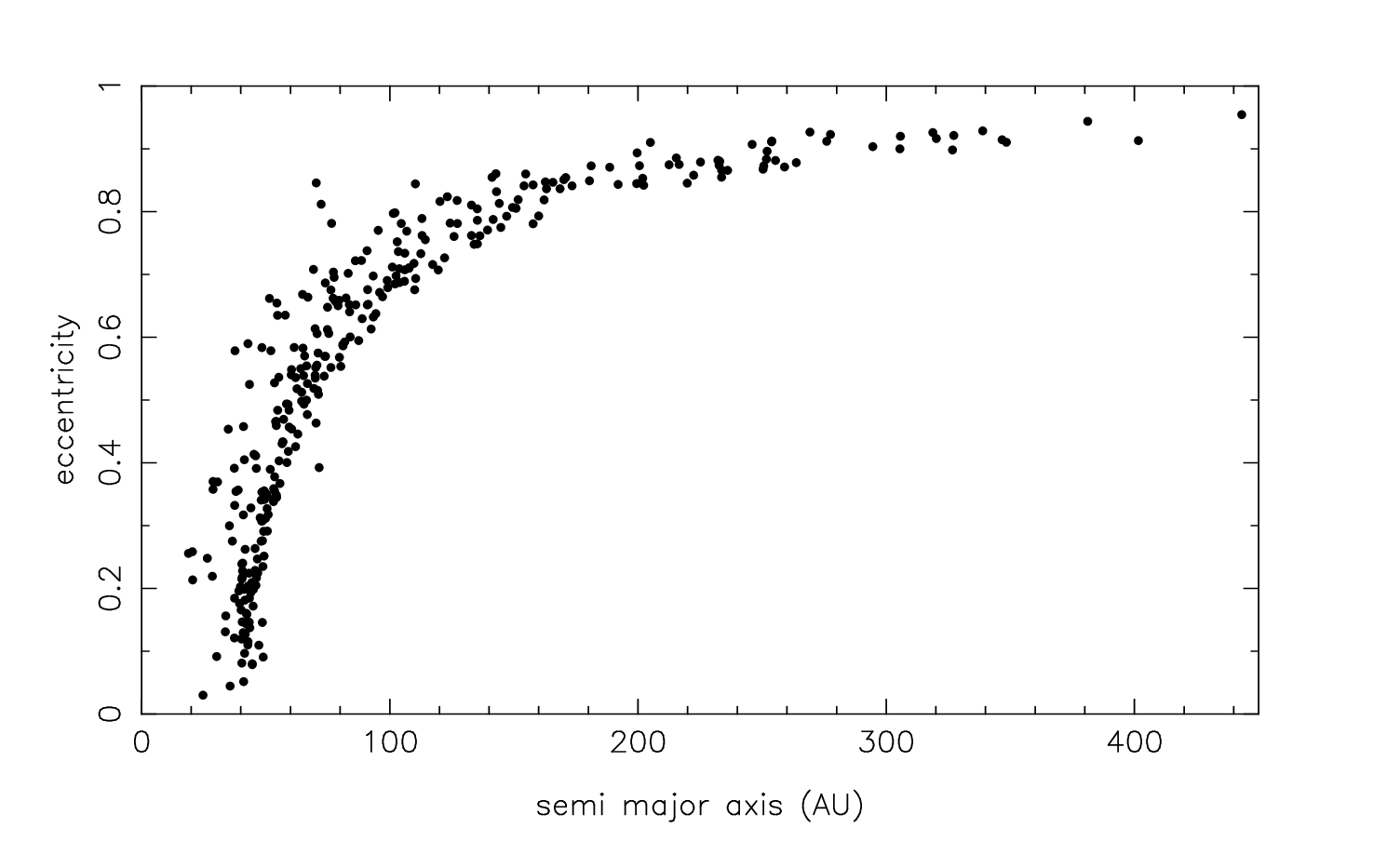}}
\caption{\small Top: the semi major axis vs. eccentricity distribution of the trans-planetary disk under the stirring effect of an embedded population of $1\,000$ Pluto-mass bodies, from Levison et al. (2011). This snapshot of the distribution is taken after 300~My of evolution. Bottom: the  semi major axis vs. eccentricity distribution of the Scattered Disk produced by the dispersal of the trans-planetary disk due to the giant planet instability, from Gomes et al. (2005). Here the snapshot of the Scattered Disk orbital distribution is taken 350~My after the beginning of the planet instability, when the Scattered Disk contains 5\% of the original disk's particles.}
\label{disk}
\end{figure}

\subsection{Numerical simulations}

In this work we use two pre-existing simulations that represent well the two phases of the evolution of the trans-planetary disk described in the Introduction: the pre-instability phase and the dynamical {dispersal} phase. 

\subsubsection{Pre-instability disk}

For the pre-instability phase we use one of the simulations of Levison et al. (2011). In those simulations, the planet instability occurs late and the disk is modeled in such a way as to mimic the self-excitation it would suffer if it contained 1,000 Pluto-mass bodies. As mentioned in the Introduction, this is a realistic number for the bodies of this reference mass in the original trans-planetary disk. We refer to Sect.~3 of Levison et al. for a technical description of how the simulation is done and to Fig.~2 of that paper for a test showing that the self-excitation process is properly reproduced. 

We took the state of the disk (\ie, the orbital distribution of the particles) 300\unit{My} after the beginning of the simulation. The self-stirring process increases the orbital excitation as $\sqrt{t}$ so that the disk is excited very rapidly, during the first few My, and then the evolution of the excitation slows down.  Thus, we take the orbital distribution of the disk at 300\unit{My} in the simulation as representative of the real dynamical state of the disk during most of its pre-instability history (we will test how the results change if the disk's state is taken at an earlier time in Sect.~\ref{pre-instability}). The top panel of Fig.~\ref{disk} shows the $(a,e)$ distribution we consider. As seen, there is a clear gradient of excitation with semi major axis. This is because (i) the shorter orbital periods in the inner part of the disk produce more frequent encounters with the massive bodies and (ii) the orbital density of the massive bodies is higher (in this simulation the initial surface density of the population of bodies in the disk is assumed to be proportional to $1/r$)\footnote{This assumption on the surface density profile is not arbitrary. In fact, an accretional proto-planetary disk with a standard $\alpha$-prescription for its viscosity (Shakura and Sunyaev, 1973) is expected to have a surface density profile proportional to $r^{-15/14}$ (Bitsch et al., 2015).}. 

 The most {up-to-date} estimate for the time of the instability, achieved by calibrating the Nice Model on various constraints (Bottke et al., 2012; Morbidelli et al., 2012; Marchi et al. 2013) is $\sim 4.1$\unit{Gy} ago, namely 450\unit{My} after the disappearance of the gas from the proto-planetary disk (4.56\unit{Gy} ago). Given the uncertainty on this estimate, and to remain conservative, we assume in the following that the pre-instability phase of the disk lasts 400\unit{My}.

\subsubsection{The dispersal of the disk and the origin of the Scattered Disk}

To study the dispersal of the trans-planetary disk and the formation of the Scattered Disk, we use one of the simulations presented in Gomes et al. (2005). In that simulation the planets become unstable at 887\unit{My} {instead of the preferred date of $\sim 450$\unit{My}. We consider only the dynamical histories of the particles after the beginning of the instability, ignoring the previous evolution, so that the actual instability date in the simulation has no influence on our results. In fact,} the dispersal of the disk in the Nice Model is a very violent event, and thus the memory of what happened in the pre-instability phase is quickly erased. Also, the overall evolution of the planetesimal population is quite insensitive to the exact evolution of the giant planets' orbits during the instability. This is evidenced by the fact that radically different simulations provide Scattered Disk populations that decay in time in very similar ways down to $\sim 1$\% of the original disk population, as reviewed in Fig.\unit{5} of Brasser and Morbidelli (2013). Thus we think that the simulation that we consider is sufficient for our purposes. 

The simulation covers a timespan of 350\unit{My} after the instability, identifies each particle individually and produces a Scattered Disk made of 5\% of the original particles at this date, whose $(a,e)$ distribution is depicted in the bottom panel of Fig.~\ref{disk}.  

\subsection{Collision probabilities and velocities}
\label{Pint}

The simulations we consider record the orbital elements of all particles at regular output intervals $dt_{out}$. Given any pair of particles, we compute their intrinsic collision probability and impact velocity using the \"Opik-like algorithm described in Wetherill (1967), implemented in a {\it fortran} code by P. Farinella and D. Davis and kindly provided to us. The algorithm considers the orbital elements $a, e, i$ (semi major axis, eccentricity and inclination) of each of the two particles and assumes that the orbital angles $\omega, \Omega, M$ (argument of perihelion, longitude of node, mean anomaly) precess linearly with time (so that their values are random on a sufficiently long time interval), without inducing changes on $(a, e, i)$. Clearly, these are approximations, but they are basically correct until the particles reach very large inclinations and undergo large amplitude Kozai cycles (Vokrouhlick\'y et al., 2012; Pokorn\'y and Vokrouhlick\'y, 2013), which is not the case for the pre-instability disk nor for most of the particles in the Scattered Disk (Kozai cycles for trans-planetary orbits are pronounced only in mean motion resonances or for planet-crossing orbits; Thomas and Morbidelli, 1996). The use of a collisional probability algorithm like Wetherill's on the output of a numerical integration is standard practice and leads to quite accurate results (\eg, Levison et al., 2000; Rickman et al., 2014). 

The code returns the {\it intrinsic collision probability} $P_i$, which is the probability that a point-like projectile hits a $R=1\unit{km}$ target in an year. Thus, the probability that two objects of radii $R_1$ and $R_2$ (expressed in km) collide over a time interval $\delta t$ is therefore $P_{coll}=P_i (R_1+R_2)^2 \delta t$.
The impact velocity $v_{coll}$ is the mean of the relative velocities between the two orbits over all collision configurations. It corresponds to the velocity of approach, prior to any acceleration due to the mutual attraction between the two bodies. The latter is negligible for planetesimals. 

For the pre-instability disk, it is not necessary to consider the collision probability of each of the simulated particles. Averaged values are enough. However, given the radial excitation gradient shown in the top panel of Fig.~\ref{disk}, we divide the disk in three zones: zone I with $15<a<20$\unit{AU}, zone II with $20<a<25$\unit{AU} and zone III with $25<a<30$\unit{AU}. Then, denoting by $k$ the particles in one zone and $m$ those in the other zone (possibly the same zones), we compute the mean intrinsic collision probability as:
\begin{equation}
\bar{P}_i={1\over{K M}}\sum_{k=1}^K\sum_{m=1}^{M} P_i(k,m)\ , 
\end{equation}
where $K$ and $M$ are the total numbers of particles in the considered disk zones and $P_i(k,m)$ is the intrinsic probability between particles $k$ and $m$. Similarly, the mean impact velocity (weighted by collision probability) is:
\begin{equation}
\bar{v}_{coll}= {1\over{KM\bar{P}_i}}\sum_{k=1}^K\sum_{m=1}^{M} P_i(k,m) v_{coll}(k,m)\ ,
\end{equation}
where $v_{coll}(k,m)$ is the collision velocity between particles $k$ and $m$.

For the simulation of the disk dispersal, instead, we compute the collision probability individually for each particle that will be a dynamical survivor in the Scattered Disk at the end of the simulation, against all other particles. Denoting by $j$ a Scattered Disk particle and by $l$ any other particle, the mean collision probability of particle $j$ at time $t$ is therefore:
\begin{equation}
\bar{P}_i(j,t)={1\over{L(t)}}\sum_{l=1}^{L(t)} P_i(j,l)\ , 
\label{Pj}
\end{equation}
where $L(t)$ is the total number of particles surviving in the integration at time $t$. For the velocity, we have:
\begin{equation}
\bar{v}_{coll}(j,t)= {1\over{L(t)\bar{P}_i(j,t)}}\sum_{l=1}^{L(t)} P_i(j,l) v_{coll}(j,l)\ .
\end{equation}

\subsection{Size distributions and disruption probabilities}
\label{absurd}

In the method introduced in CM03/07, one defines an initial size distribution for the swarm of planetesimals represented by each simulation particle. At each step, using the pre-computed collision probabilities among the simulation particles, one computes the number of collisions occurring between planetesimals of any given size. From the impact velocities and masses of projectile and target, one computes the consequences of the collisions (cratering event, catastrophic break-up) and the size distribution of the generated fragments. In this way, the evolution of the size distributions associated to each simulation particle is computed. In the end, one evaluates which fraction of the planetesimals of a given size are survivors of the original population or collisional fragments of larger planetesimals. 

This approach is correct, but it is overshooting for our goal in this paper, which is just to assess whether a comet-size object is likely to have avoided catastrophic collisions. Moreover, it requires exploring a variety of initial size distributions, demanding a quite tedious exploration of the parameter space defining them. 

Thus, we modify and simplify the approach as described below. We use the approach typical of a mathematical demonstration {\it ad absurdum} (by reduction to the absurd). That is, we start by {\it assuming} that the planetesimals down to comet-size objects are not significantly affected by collisions. This means that the planetesimal size distribution does not evolve with time, and that the initial distribution has to be the same as the current distribution in the Scattered Disk, just scaled up by the inverse of the implantation efficiency (the fraction of the disk population surviving in the end in the Scattered Disk). We detail below what this size distribution is.

Then, using this distribution and the pre-computed collision probabilities and velocities, we evaluate the minimum size of a projectile capable of disrupting a comet-size body and thus the number of catastrophic collisions $n_{coll}$ that each comet-size body should suffer (we detail below how we do this evaluation). The probability that a comet-size body has escaped all catastrophic collisions is then $P_{intact}=\exp(-n_{coll})$. If $P_{intact}$ is close to unity, then our assumption of a negligible collisional role is verified. But if $P_{intact}$  is small, then we reach the absurd situation that, by assuming that the planetesimal population was not affected by collisions, we conclude that most planetesimals should have been destroyed! This means that the assumption was wrong, and hence, that the planetesimal size distribution was significantly affected by collisions.

With this approach we cannot compute the actual probability that a comet-like body has escaped all catastrophic collisions, but we know that it has to be smaller than $P_{intact}$. In fact, any initial size distribution evolving by collisions towards the current Scattered Disk distribution must have had originally more bodies than we assumed (because of collisional comminution) and therefore, the probability that a given body was catastrophically disrupted must be higher than we computed (because of a larger initial number of projectiles). If the value of $P_{intact}$  that we computed is already small, this is enough for our purposes. 

In this paper, as comet-size bodies we consider objects with radius $R=2$\unit{km}, appropriate for comet 67P/Churyumov-Gerasimenko, the target of the Rosetta mission (see Rickman et al., 2015).

\subsubsection{The Scattered Disk size distribution and the minimal number of comet-size objects in the original trans-planetary disk}\label{SD}

The most recent estimate of the Scattered Disk population has been presented in Brasser and Morbidelli (2013). In that work, as in Duncan and Levison (1997), the number of comet-size bodies in the Scattered Disk is evaluated from (a) the number of known Jupiter family comets in some given range of orbits and magnitudes for which the JFC sample is assumed complete and (b) the numerical relationship between the Scattered Disk population and the Jupiter family population that the former sustains, obtained from numerical simulations.  With respect to previous estimates (\eg, Duncan and Levison, 1997), the estimate in Brasser and Morbidelli is improved in two respects: it uses the most recent conversion from total magnitude to nuclear size from Fern\'andez and Sosa (2012) and it is based on new simulations deriving  Jupiter-family comets from a Scattered Disk that is excited in inclination (the original Duncan and Levison work assumed that inclinations in the Scattered Disk are of a few degrees only, which has then been refuted by observations). Brasser and Morbidelli concluded that there are $2\times 10^9$ bodies in the Scattered Disk today larger than 2.3\unit{km} in diameter. Given a Scattered Disk implantation efficiency of 1\%, this means that the original trans-planetary disk, if not affected by collisional comminution, should have contained $2\times 10^{11}$ of these bodies. 

We believe that this is a lower bound for the original disk population for three reasons. First, in a subsequent work accounting for an improved fading law (probability that a comet does not survive more than $n$ perihelion passages), Brasser and Wang (2014) raised the estimate of the Scattered Disk population to $6\times 10^9$ bodies larger than 2.3\unit{km} in diameter. Second, serendipitous stellar occultation observations by the HST guiding sensors (Schlichting et al., 2009) and by the Corot survey (Liu et al., 2015) suggest that the average sky density of bodies larger than 250\unit{m} in radius over a $\pm 5^\circ$ ecliptic band is $2\times 10^7$/deg$^2$. Thus, there are at least $7\times 10^{10}$ bodies of this size in the trans-Neptunian region; with a cumulative size distribution proportional to $R^{-2}$ this implies $3.5\times 10^{9}$ objects with $D>2.3$\unit{km}. It is unclear which population (cold EKB, hot EKB, Scattered Disk) the detected objects belong to. But, given that the Scattered Disk outnumbers the others (compare Trujillo et al., 2000 with Fraser et al., 2014 for the observational point of view and Brasser and Morbidelli 2013 with Nesvorn\'y, 2015 for the modeling point of view), the number above can be considered to be an estimate -- if not a lower bound -- of the Scattered Disk population. Third, repeating the same exercise for the Oort cloud population, Brasser and Morbidelli (2013) estimated that the primordial trans-planetary disk should have contained $10^{12}$ objects with $D>2.3$\unit{km}; thus probably the reality lies in between $2\times 10^{11}$ and $10^{12}$. 

Thus, to remain conservative (\ie, underestimate the total number of collisions), in this work we assume that the trans-planetary disk contained $2\times 10^{11}$ objects with $D>2.3$\unit{km}. As for the size distribution, we turn again to comet observations. Estimates of the JFC size distribution range significantly from authors to authors, from quite steep (exponent of the differential distribution {close} to $-3.5$ -- Fern\'andez et al., 1999; Tancredi et al., 2006 -- or even steeper -- Belton, 2015) to shallow (differential slope of $-2.6$;  Lowry and Weissman, 2003). Consequently, in this work we assume as nominal differential slope the value $-3$ (Meech et al., 2004; Lamy et al., 2004; Snodgrass et al., 2011), but we also study the dependence of the results on exponents for the {differential} distribution ranging from $-2.5$ to $-3.5$. 

{A shallow size distribution is preferred according the the most recent planetesimal formation models (Johansen et al., 2015: $q=-2.8$). TNO surveys (e.g. Fraser et al., 2014) also suggest that the size distribution of objects smaller than 50\unit{km} in radius is shallow ($q$ between $-3.1$ and $-2.5$, although it may steepen up for not yet detectable comet-size bodies).}

For reference, a disk with a size distribution similar to that of the hot EKB (Fraser et al., 2014), namely with a differential slope of $-3$ for $R<50$~\unit{km} and $-5$ for $R>50$~\unit{km}, a total number of $2\times 10^{11}$ objects with $R>1.15$~\unit{km} and a density of 1\unit{g}/\unit{cm}$^3$ would have a total mass of 35~Earth masses, in good agreement with that required by the Nice Model (Gomes et al, 2005; Morbidelli et al., 2007; Nesvorn\'y and Morbidelli, 2012).  

\subsubsection{Minimum size of catastrophic projectiles}
\label{size}

The kinetic energy of an impact that can catastrophically destroy an object is 
\begin{equation}
E_{disrupt}={4\over 3} \pi \rho R^3 Q^*(R)\ ,
\label{Edisrupt}
\end{equation}
where $\rho$ is the bulk density of the target of radius $R$; $Q^*(R)$ is the specific energy for disruption and is size dependent. There are several $Q^*(R)$ laws proposed in the literature for various materials. Benz and Asphaug (1999) propose two such laws for bodies made of "strong ice", hit respectively at 1\unit{km/s} and 3\unit{km/s}. As we will see in Sect.~\ref{results}, the former velocity is well adapted to the pre-instability disk while the second is suitable for the disk dispersal phase.  For a $R=2$\unit{km} body the two values of $Q^*(R)$ are actually comparable. Leinhardt and Stewart (2009) produced a $Q^*$ law for bodies made of "weak ice", hit at 1\unit{km/s}. Their $Q^*(R)$ function follows the general trend of those in Benz and Asphaug (1999) but the value for $R=2$\unit{km} is about an order of magnitude smaller (see their Fig. 11). The scaling of $Q^*$ with velocity given in Eq.~(2) of Stewart and Leinhardt (2009) gives a value 2.4 times larger for a velocity of 3\unit{km}/\unit{s}, which is still 4 times smaller than that reported in  Benz and Asphaug (1999) for the same speed.
Even if the Leinhardt and Stewart value may be more appropriate for pristine, low-density planetesimals, we use the Benz and Asphaug values in order to be conservative once again. This likely overestimates the minimal size of a projectile capable of disrupting the target and thus underestimates the number of catastrophic collisions. 

In fact, once $E_{disrupt}$ is known, the minimal size of a catastrophic projectile $r_p$ is given by the equation
\begin{equation}
{4\over 3} \pi \rho r_p^3 {1\over 2} v_{coll}^2=E_{disrupt} \ ,\label{proj-radius}
\end{equation}  
so that, the higher is $E_{disrupt}$ the larger is $r_p$. Notice that, if one assumes that the bulk density of projectile and target is the same, $\rho$ simplifies from the right hand and left hand sides of (\ref{proj-radius}) and the result is independent of $\rho$.

\subsubsection{Total number of catastrophic events}

Once the minimum size of a catastrophic projectile is known, the total number of catastrophic impacts for a target of radius $R_T$ is computed as:
\begin{equation}
N_{coll}=(\bar{P}_i \delta t)\int_{r_p}^{R_{max}} (R_T+R_p)^2 N(R_p) dR_p \ ,
\label{Ncoll}
\end{equation}
where $\bar{P}_i$ is the considered intrinsic probability (averaged over the ensemble of potential projectiles, as explained in Sect.~\ref{Pint}), $\delta t$ is the considered time-span, $N(R_p)dR_p$ is the differential size distribution, $r_p$ is the minimum size for a catastrophic projectile and $R_{max}$ is the maximum size for which the considered size distribution is valid. Given that the size distribution of the trans-Neptunian populations turns from steep (at the large size end) to shallow (at the small size end) at a size of approximately $R\sim 50$\unit{km} (Bernstein et al., 2004; Fuentes et al., 2009; Fraser et al., 2014), we assume $R_{max}=50$\unit{km}. We neglect the relatively small contribution by projectiles of even larger sizes.

Eq.~(\ref{Ncoll}) is often approximated by 
\begin{equation}
N_{coll}=\bar{P}_i \delta t R_T^2 \int_{r_p}^{R_{max}} N(R_p) dR_p  =\bar{P}_i \delta t R_T^2 N(>r_p) \ ,
\label{approx}
\end{equation}
where $N(>r_p)$ is the cumulative number of bodies larger than $r_p$. This approximation is good for steep size distributions, or in the limit $r_p\to 0$. However, for shallow size distributions like the one we consider here and $r_p$ not much smaller than $R_T$ (as is the case for low velocity collisions), the approximation is not very precise. Thus, we solve the integral (\ref{Ncoll}) exactly. That is, denoting by $q$ the exponent of the differential size distribution, the primitive of the integrand in (\ref{Ncoll}) is:
\begin{equation}
\begin{array}{lcl}
q&=&-2.5:\quad -2 (R_T^2 + 6 R_T R_p -3 R_p^2)/(3 R_p^{3/2})\\
q&=&-3:\quad -R_T^2/(2 R_p^2) - 2 R_T/R_p + \log(R_p) \\
q&=&-3.5:\quad -2 (3 R_T^2 +10 R_T R_p + 15 R_p^2)/(15 R_p^{5/2})\ .
\label{primitives}
\end{array}
\end{equation} 

\section{Results}
\label{results}

We report here the results obtained for the pre-instability disk and the disk dispersal phase, obtained applying the methods described in the previous Section. 

\subsection{Disruptive collisions in the pre-instability disk}
\label{pre-instability}

We show in Table~\ref{disk-Pvr} the results concerning the intrinsic collision probability $\bar{P}_i$, the collision velocity $v_{coll}$ and the minimum size of a catastrophic projectile $r_p$ for a target with $R=2$\unit{km}, considering all possible combinations of disk zones for projectile and target.

{\small 
\begin{table}[t]
\caption{Table of results for the pre-instability disk. The first row reports the disk zone where the target is located and the first column reports the disk zone where the projectile is located. The disk zones are: (I) $a<20$\unit{AU}, (II) $20<a<25$\unit{AU}, (III) $a>25$\unit{AU}. Then, each box reports on the top the mean intrinsic collision probability $\bar{P}_i$ (number of collisions per year per projectile on a target of $R=1$\unit{km}), in the middle the mean collision velocity $v_{coll}$ (in km/s) and at the bottom the minimum size of a catastrophic projectile $r_p$ (in km).}
\vspace{.1cm}
\begin{center}
\begin{tabular}{|c|c|c|c|} \\ 
$_{\rm projectile} \backslash ^{\rm target}$ & I  & II & III \\
\hline
\multirow{4}{1pt}{I}& & & \\
                  & $1.85\times 10^{-20}$ & $3.75\times 10^{-21}$ & $1.00\times 10^{-24}$ \\ 
                  & 0.78 & 0.74 & 0.95 \\ 
                  & 0.23 & 0.24 & 0.20 \\ \hline
\multirow{4}{1pt}{II}& & & \\
              & $3.75\times 10^{-21}$ & $8.95\times 10^{-21}$ & $7.95\times 10^{-22}$ \\ 
                  & 0.74 & 0.44 & 0.38 \\ 
                  & 0.24 & 0.33 & 0.37 \\ \hline
\multirow{4}{1pt}{III}& & & \\
                 & $1.00\times 10^{-24}$ & $7.95\times 10^{-22}$ & $7.32\times 10^{-21}$ \\ 
                    & 0.95 & 0.38 & 0.24 \\ 
                    & 0.20 & 0.37 & 0.51 \\ \hline
\end{tabular}
\end{center}
\label{disk-Pvr}
\end{table}
}

As seen, the intrinsic collision probability is higher if both target and projectile are in the inner part of the disk than in the outer part. The collision velocity is also higher. The mean intrinsic collision probability is lower if target and projectile belong to different disk zones, because not all particle orbits from the two zones intersect. 

Table~\ref{TN} reports the number of catastrophic collisions expected for a 2\unit{km} target, for the three considered values of the exponent $q$ of the differential size distribution. The calculation has been done assuming $\delta t=400$\unit{My} (the expected lifetime of the pre-instability phase), and applying (\ref{Ncoll}) and (\ref{primitives}) to the numbers reported in Table~\ref{disk-Pvr} for projectiles in each disk zone. Because the surface density of the disk is assumed to be proportional to $1/r$ in the Levison et al. (2011) simulation that we use, an equal number of projectiles of a given size is assumed to exist initially in each of the disk zones.

{\small 
\begin{table}[t]
\caption{Number of disruptive collisions expected for a $R=2$\unit{km} target located in each disk zone, as a function of the exponent $q$ of the differential size distribution. The first row reports the target's zone. The first column gives the value of $q$. Each box reports the number of catastrophic collisions expected over 400\unit{My}. In {parentheses} we report the same quantity estimated by using the dynamical state of the disk after 100\unit{My} of evolution, instead of that shown in the top panel of Fig.~\ref{disk} (300\unit{My}). The number of catastrophic collisions is smaller, but it is nevertheless much larger than unity in all cases.}
\vspace{.1cm}
\begin{center}
\begin{tabular}{c|c|c|c|}
$q \backslash ^{\rm target zone}$ & I  & II & III \\
\hline
$-2.5$& $58.0$ ($51.2$) &  $28.7$ ($20.7$) & $12.3$ ($9.6$)\\
\hline
$-3.0$& $94.5$ ($75.0$) & $39.7$ ($23.7$) & $12.1$ ($7.9$) \\
\hline
$-3.5$& $190.6$ ($137.7$) & $70.2$ ($35.3$) & $15.4$ ($8.2$)\\
\hline
\end{tabular}
\end{center}
\label{TN}
\end{table}
}

{\small 
\begin{table}[t]
\caption{The same as Table~\ref{TN} but now reporting the radius of a body (in \unit{km}) for which the probability of catastrophic impact is 50\%, for each disk zone and assumed slope of the projectile size distribution. In {parentheses} we report the same quantity estimated by using the dynamical state of the disk after 100\unit{My} of evolution, instead of that shown in the top panel of Fig.~\ref{disk} (300\unit{My}). When the result exceeds 50\unit{km} (the value at which the size distribution is no longer described by a power-law with exponent $q$), we report 50 for simplicity.}
\vspace{.1cm}
\begin{center}
\begin{tabular}{c|c|c|c|}
$q \backslash ^{\rm target zone}$ & I  & II & III \\
\hline
$-2.5$& $50$ ($50$) &  $50$ ($50$) & $50$ ($43$)\\
\hline
$-3.0$& $50$ ($50$) & $50$ ($50$) & $48$ ($37$) \\
\hline
$-3.5$& $50$ ($50$) & $50$ ($44$) & $29$ ($20$)\\
\hline
\end{tabular}
\end{center}
\label{sizemax}
\end{table}
}

We notice that the total number of collisions for comets in zone III of the disk has a very weak dependence on $q$, because the size of the minimum catastrophic projectile is quite big (0.5\unit{km} in radius). The opposite is true for comet-size targets in zone I of the disk. Clearly, in all cases the total number of collisions is larger than 1. This means that the probability that a 2\unit{km} body escapes from all catastrophic collisions is small. From the numbers in Table~\ref{TN} this probability is always smaller than $\exp(-12)=6\times 10^{-6}$; using the numbers in parentheses, we get $\exp(-7.9)=4\times 10^{-4}$.

{Table~\ref{sizemax} reports the radius of the comets for which the probability of a catastrophic impact over the age of the disk is 50\%, again for each disk zone and assumed value of $q$. This size is extremely large, in most cases exceeding 50\unit{km}. This is beacuse the number of catastrophic impacts depends weakly on the size of the target. In fact, if one assumes for simplicity that $Q^*$ is independent of size, the size of the minimal catastrophic projectile scales linearly with the target size $R_T$; then, using (\ref{approx}) one finds that the number of catastrophic impacts scales as $R_T^{2+q+1}$ which, for $q=-3$ eliminates the dependence on $R_T$. For the compilation of Table~\ref{sizemax} we nevertheless used the dependence of $Q^*$ on radius given in Benz and Asphaug (1999) and the non-approximated formul\ae\ (\ref{primitives}).}

Note that this result is valid for both Scattered Disk comets (JFCs) and Oort cloud comets (LPCs/HTCs), because both reservoirs form from the same disk (Brasser and Morbidelli, 2013). Thus we conclude that if the giant planet instability occurred as late as 4.1\unit{Gy} ago, the possibility that a 2\unit{km} comet is a pristine planetesimal, rather than a collisional fragment, is very slim.

\subsection{Disruptive collisions during disk dispersal}
\label{dispersal}

As we said in Sect.~\ref{Pint}, for the disk dispersal phase  we consider individually each particle ending in the Scattered Disk, because their orbital histories can be very diverse. However, because the assumption in our {\it ad absurdum} approach is that the disk is not collisionally active and its size distribution does not evolve, for each target particle $j$ we can average over time the value of $\bar{P}_i(j,t)$ given in (\ref{Pj}) as:
\begin{equation}
\bar{\bar{P}}_i(j)={1\over {L(0)}} \sum_t L(t) \bar{P}_i(j,t)
\end{equation}
and apply the result over a time interval $\delta t=350$\unit{My}, which is the integration timespan.

However, this 350\unit{My} simulation timespan covers only a small fraction of the lifetime of the Solar System and in principle there may still be a significant collisional evolution in the Scattered Disk over the remaining $\sim 4$\unit{Gy} of Solar System history. To have an estimate of the collision probability over this remaining time, we proceed as follows. We assume that the orbital distribution in the Scattered Disk does not evolve with time, remaining equal to that shown in the bottom panel of Fig.~\ref{disk}, but we assume that the Scattered Disk population decays with time. The Scattered Disk at the end of the simulation accounts for 5\% of the initial disk particles and over the rest of the Solar System lifetime it would decay to about 1\% (Brasser and Morbidelli, 2013).  So, we assume that number of Scattered Disk particles decays as $\exp(-t/\tau)$ with $\tau$ such that after 4\unit{Gy} the population is reduced by a factor of 5. The integral of the exponential function over 4\unit{Gy} with such a value of $\tau$ is $0.5$. Thus, we take the last computed value of $\bar{P}(j,t)$ for each Scattered Disk particle ($t=350$\unit{My}) and we multiply it by $\delta t=4$\unit{Gy} and then divide by 2. We find that the integrated collision probability over the last 4\unit{Gy} is a fraction (typically from 10\% to 80\%) of that integrated over the first 350\unit{My}. This is because at the beginning the disk is much more populated and, moreover, the most collisional active phase is when the disk is just stirred up by the planets' action (Stern and Weissman, 2001). Instead, once the Scattered Disk is formed, the collision probability per unit time per particle is strongly reduced, due to the large orbital space that the Scattered Disk fills and the long orbital periods. 

Given the typical collision velocities of $2-4$\unit{km}/\unit{s}, the typical size of the smallest catastrophic projectile for a $R=2$\unit{km} target is $\sim 100$\unit{m}. One may wonder whether bodies this small existed in the disk. The occultation observations mentioned above (Schlichting et al., 2009; Liu et al., 2015) show that bodies of this size exist today. Following our {\it ad absurdum} approach, we then need to assume that they existed in the original disk, because if the disk did not evolve collisionally, no objects could be generated. 

\begin{figure}[t!]
\centerline{\includegraphics[height=5.5cm]{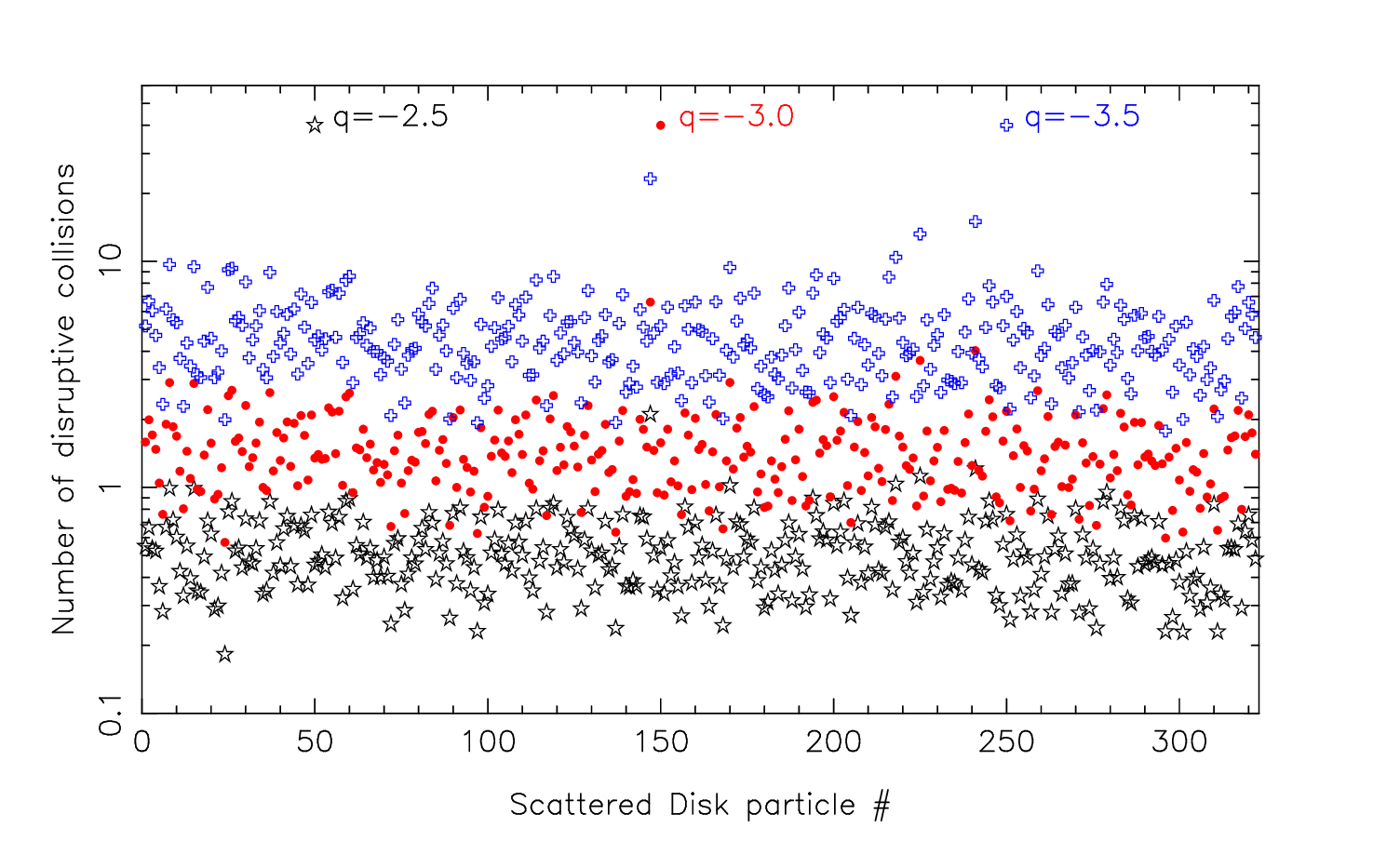}}
\caption{\small The number of expected catastrophic collision for each particle surviving in the Scattered Disk at the end of the disk dispersal simulation. The symbols depict different values for the exponent of the differential size distribution $q$, as labeled in the plot.}
\label{final-dispersal}
\end{figure}

Fig.~\ref{final-dispersal} shows for each Scattered Disk particle the expected number of catastrophic collisions that a 2\unit{km} target should have suffered; the colors refer to different values of $q$.  The number of catastrophic collisions changes considerably from particle to particle, because the dynamical histories of Scattered Disk objects can be very diverse. We see that, if $q=-3.5$ or steeper, clearly each comet-size object should have suffered at least two catastrophic collisions, with an average of 4.7 collisions. {The radius of comets for which the average number of catastrophic impacts would be 0.5 is approximately 10\unit{km}.} This excludes the possibility, proposed by Belton (2015), that the size distribution of the Scattered Disk is steeper than $q=-3.5$ {below this radius}. In fact, the disk would be collisionally evolved and therefore it would have acquired a collisional equilibrium size distribution, which implies $q=-3.5$ (Dohnanyi, 1969) or shallower ($|q|<3.5$; O'Brien and Greenberg, 2003). 

If the distribution is very shallow ($q=-2.5$), Fig.~\ref{final-dispersal} shows that about half of the comets should have had no catastrophic collisions, the average number of catastrophic collisions per object being 0.5. The nominal case $q=-3.0$ is borderline. Most comets should have had at least one catastrophic collision (the average being 1.5 collisions per comet) but some, with favorable orbital histories, would have had no collision at all. Please notice that, had we used the 4 times smaller value of $Q^*$ from Leinhardt and Stewart (see Sect.~\ref{size}), the number of catastrophic impacts would have increased by a factor of $\sim 3$ for $q=-3.5$, a factor $\sim 2.5$ for $q=-3$ and a factor of 2 for $q=-2.5$. 

\begin{figure}[h!]
\centerline{\includegraphics[width=6.cm,angle=-90]{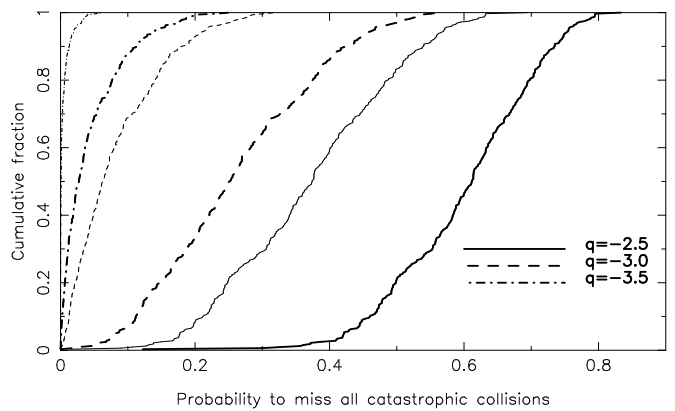}}
\caption{\small Fraction of particles ending in the Scattered Disk with a probability to escape all catastrophic collisions $P(0)$ smaller than that indicated on the horizontal axis. This is an alternative represenation of the results already shown in Fig.~\ref{final-dispersal}. The different line styles refer to different exponents for the differential size distribution $q$, as labeled on the plot. The thick curves correspond to the value of $Q^*$ given in Benz and Asphaug (1999) and the thin curves to a $Q^*$ value 4 times smaller, as in Leinhardt and Stewart (2009) with the velocity scaling provided in Stewart and Leinhardt (2009).}
\label{Cumul}
\end{figure}

Fig.~\ref{Cumul} shows the same results but using a different representation. The number of collisions $N_{coll}$ is converted into a probability to avoid all collisions $P(0)=\exp(-N_{coll})$. Then, the normalized cumulative distribution of the $P(0)$ values is plotted. The thick curves are for the nominal $Q^*$ value from Benz and Asphaug (1999) and the thin curves for a $Q^*$ value 4 times smaller.  

Note that, if the disk dispersal occurred late, the number of catastrophic collisions found in this Section should be added to those reported in Table~\ref{TN}. So, all original comets should have been destroyed. If instead the dispersal of the disk occurred soon after the disappearance of gas from the disk, the results illustrated in Figs.~\ref{final-dispersal} and~\ref{Cumul} apply alone. In this case, if the original and current size distributions in the comet-size range are shallow and a quite large value of $Q^*$ applies, there is the possibility that a fraction of the comets are pristine objects which escaped catastrophic collisions.

\section{Conclusions}
\label{conclusions}

In this work we have estimated the total number of catastrophic collisions that a typical Jupiter family comet (here assumed to have radius $R=2$\unit{km}) should have had over its dynamical lifetime, first in the trans-planetary disk and then in the Scattered Disk. 

We have shown that, if the trans-planetary disk beyond the original orbit of Neptune has been dispersed by a late dynamical instability of the giant planets occurring $\sim 4.1$\unit{Gy} ago, comet-size objects should have suffered numerous catastrophic collisions in the pre-instability phase. Thus, not only JFCs, but also Oort cloud comets should be fragments of originally larger bodies. Because the late instability of the giant planet system is, at the current level of understanding, the best explanation for the trigger of the Late Heavy Bombardment of the Solar System, the formation of late lunar basins on the Moon (Bottke et al., 2012; Morbidelli et al., 2012) and the impact age record on meteorites (Marchi et al., 2013), we believe that this is the conclusion of our work.   

However, in the hypothetical case that the dispersal of the disk occurred early, the collisional evolution of comet-size bodies ending in the Scattered Disk would have been less severe. If the size distribution of comet-size objects in today's Scattered Disk and in the primordial trans-planetary disk was shallow (differential index $|q|\lta 3$), it is possible in principle that a significant fraction of comet-size objects escaped all catastrophic collisions.

The reader should keep in mind, though, that throughout our study we have taken the most conservative assumptions, so that the number of catastrophic collisions that we computed should be considered as a lower estimate. In fact, we have considered an initial size distribution in the disk that contains the minimum possible number of comet-size objects (Sect.~\ref{SD}). Also, we have assumed the specific energy for catastrophic disruption given in Benz and Asphaug (1999), which probably overestimates that appropriate for weak icy aggregates (Leinhardt and Stewart, 2009; Sect.~\ref{size}): the adoption of the 4 times smaller specific energy for disruption of Leinhardt and Stewart (2009) would have multiplied the number of catastrophic impacts shown in Fig.~\ref{final-dispersal} by a factor 2.5 for $q=-3$ and a factor of 2 if $q=-2.5$, while producing the thin curves in Fig.~\ref{Cumul}. Moreover, the {\it ad absurdum} approach that we followed by its essence provides just a {\it minimal} estimate of the number of collisions (Sect.~\ref{absurd}). Finally, one should take into account that, for each catastrophic collision, the number of quasi-catastrophic collisions would be much higher (being caused by smaller projectiles, which are more numerous). Thus, even if a comet had not suffered, by chance, any catastrophic collision, its morphology would have been sculpted but numerous large sub-catastrophic impacts.

Therefore, we conclude that typical JFCs of the size of 67P/Churyumov-Gerasimenko most likely are not intact planetesimals but they are either fragments of originally bigger bodies (the most likely case), or are planetesimals strongly sculpted by barely sub-catastrophic impacts\footnote{The special case of comet 67P and constraints on its origin has been treated by Rickman et al. (2015).}. In the latter case, this sculpting may disappear after the comets have been severely eroded by sublimation or splitting. However, some JFCs are statistically likely to bear these scars.

There are some properties of comets that appear to be in contradiction to a collisional origin and would rather suggest that comets are primordial survivors. In the Introduction we mentioned the low bulk densities and negligible tensile strengths along with the large abundance of super-volatiles like CO. These properties are certainly compatible with an origin of comets as primordial survivors of the icy planetesimal population. Hence, our results lead to the obvious question of whether collisions are able to conserve these primordial properties in the fragments they produce. This question is currently unanswered and merits careful consideration.

{If a detailed modeling of the collisions between icy planetesimals would show that the primordial-like features of comets are not preserved in the fragments, one may suspect that our current vision of outer Solar System evolution is not appropriate. For instance, there might have been no delay in the dynamical instability, or the disk remained less self-excited due to a smaller number of large bodies than we envision, or there was a drastic cut-off in the size distribution affecting sub-km objects, thus limiting the number of catastrophic projectiles.}

\section{Acknowledgments}

A.M. was supported by ANR, pro
ject number ANR-13--13-BS05-0003-01  projet MOJO(Modeling the Origin of JOvian p
lanets). H.R. was supported by Grant No. 2011/01/B/ST9/05442 of the Polish National Science Center. {The authors thank an anonymous reviewer for constructive comments.}

\section{References}

\begin{itemize}

\item A'Hearn, M.~F., Feaga, L.~M., Keller, H.~U., et al.\ 2012, \apj, 758, 29 
\item Altwegg, K., Balsiger, H., Bar-Nun, A., et al.\ 2015, Science, 347, A1261952 
\item Asphaug, E., \& Benz, W.\ 1996, \icarus, 121, 225 
\item Belton, M.~J.~S.\ 2015, \icarus, 245, 87 
\item Benz, W., \& Asphaug, E.\ 1999, \icarus, 142, 5
\item Bernstein, G.~M., Trilling, D.~E., Allen, R.~L., et al.\ 2004, \aj, 128, 1364 
\item Bitsch, B., Johansen, A., Lambrechts, M., \& Morbidelli, A.\ 2015, \aap, 575, AA28 
\item {Blum, J., Gundlach, B., 
M{\"u}hle, S., \& Trigo-Rodriguez, J.~M.\ 2014, \icarus, 235, 156} 
\item Bockel\'ee-Morvan, D., Crovisier, J., Mumma, M.~J., \& Weaver, H.~F.\ 2004, Comets II, 391
\item Bottke, W.~F., Levison, H.~F., Nesvorn{\'y}, D., \& Dones, L.\ 2007, \icarus, 190, 203 
\item Bottke, W.~F., Vokrouhlick{\'y}, D., Minton, D., et al.\ 2012, \nat, 485, 78 
\item Brasser, R., \& Morbidelli, A.\ 2011, \aap, 535, AA41 
\item Brasser, R., \& Morbidelli, A.\ 2013, \icarus, 225, 40 
\item Brasser, R., \& Wang, J.-H.\ 2015, \aap, 573, AA102 
\item Chambers, J.~E.\ 2007, \icarus, 189, 386 
\item Charnoz, S., \& Morbidelli, A.\ 2003, \icarus, 166, 141 
\item Charnoz, S., \& Morbidelli, A.\ 2007, \icarus, 188, 468 
\item {\'C}uk, M., \& Gladman, B.~J.\ 2006, \icarus, 183, 362 
\item {\'C}uk, M., \& Gladman, B.~J.\ 2009, \icarus, 199, 237 
\item {\'C}uk, M.\ 2012, \icarus,  218, 69 
\item Davidsson, B.~J.~R., Guti{\'e}rrez, P.~J., \& Rickman, H.\ 2007, \icarus, 187, 306 
\item Davis, D.~R., \& Farinella, P.\ 1997, \icarus, 125, 50 
\item Dohnanyi, J.~S.\ 1969, \jgr, 74, 2531 
\item Duncan, M.~J., \& Levison, H.~F.\ 1997, Science, 276, 1670 
\item Fern{\'a}ndez, J.~A., Tancredi, G., Rickman, H., \& Licandro, J.\ 1999, \aap, 352, 327 
\item Fern{\'a}ndez, J.~A., \& Sosa, A.\ 2012, \mnras, 423, 1674 
\item Fraser, W.~C., Brown, M.~E., Morbidelli, A., Parker, A., \& Batygin, K.\ 2014, \apj, 782, 100 
\item Fuentes, C.~I., George, M.~R., \& Holman, M.~J.\ 2009, \apj, 696, 91 
\item Gomes, R., Levison, H.~F., Tsiganis, K., \& Morbidelli, A.\ 2005, \nat, 435, 466 
\item Lamy, P.~L., Toth, I., Fernandez, Y.~R., \& Weaver, H.~A.\ 2004, Comets II, 223 
\item Leinhardt, Z.~M., \& Stewart, S.~T.\ 2009, \icarus, 199, 542 
\item Levison, H.~F., Duncan, M.~J., Zahnle, K., Holman, M., \& Dones, L.\ 2000, \icarus, 143, 415 
\item Levison, H.~F., Morbidelli, A., Tsiganis, K., Nesvorn{\'y}, D., \& Gomes, R.\ 2011, \aj, 142, 152 
\item Liu et al., 2015. MNRAS, in press.
\item Lowry, S.~C., \& Weissman, P.~R.\ 2003, \icarus, 164, 492 
\item Luu, J., Marsden, B.~G., Jewitt, D., et al.\ 1997, \nat, 387, 573 
\item Marchi, S., Bottke, W.~F., Cohen, B.~A., et al.\ 2013, Nature Geoscience, 6, 303 
\item Meech, K.~J., Hainaut, O.~R., \& Marsden, B.~G.\ 2004, \icarus, 170, 463 
\item Minton, D.A., Jackson, A.P., Asphaug, E., Fassett, C.I., Richardson, J.E., 2015. Early Solar System Bombardment III, abs. \#3033
\item Morbidelli, A., Tsiganis, K., Crida, A., Levison, H.~F., \& Gomes, R.\ 2007, \aj, 134, 1790 
\item Morbidelli, A., Marchi, S., Bottke, W.~F., \& Kring, D.~A.\ 2012, Earth and Planetary Science Letters, 355, 144 
\item Nesvorn{\'y}, D., \& Morbidelli, A.\ 2012, \aj, 144, 117 
\item Nesvorn\'y, D., 2015, submitted.
\item O'Brien, D.~P., \& Greenberg, R.\ 2003, \icarus, 164, 334 
\item Pokorn{\'y}, P., \& Vokrouhlick{\'y}, D.\ 2013, \icarus, 226, 682 
\item Rickman, H.\ 1989, Advances in Space Research, 9, 59 
\item Rickman, H.\ 2004, Comets II, 205 
\item Rickman, H., Wi{\'s}niowski, T., Wajer, P., Gabryszewski, R., \& Valsecchi, G.~B.\ 2014, \aap, 569, AA47 
\item Rickman, H., Marchi, S., A'Hearn, M.~F., et al.\ 2015, \aap, submitted.
\item Robuchon, G., Nimmo, F., Roberts, J., \& Kirchoff, M.\ 2011, \icarus, 214, 82 
\item Safronov, V.~S.\ 1977, IAU Colloq.~39: Comets, Asteroids, Meteorites: Interrelations, Evolution and Origins, 483 
\item Schlichting, H.~E., Ofek, E.~O., Wenz, M., et al.\ 2009, \nat, 462, 895
\item Schlichting, H.~E., Fuentes, C.~I., \& Trilling, D.~E.\ 2013, \aj, 146, 36 
\item Shakura, N.~I., \& Sunyaev, R.~A.\ 1973, \aap, 24, 337 
\item Snodgrass, C., Fitzsimmons, A., Lowry, S.~C., \& Weissman, P.\ 2011, \mnras, 414, 458 
\item Stern, S.~A.\ 1991, \icarus, 90, 271 
\item Stern, S.~A., \& Weissman, P.~R.\ 2001, \nat, 409, 589 
\item Stewart, S.~T., \& Leinhardt, Z.~M.\ 2009, \apjl, 691, L133 
\item Tancredi, G., Fern{\'a}ndez, J.~A., Rickman, H., \& Licandro, J.\ 2006, \icarus, 182, 527 
\item Thomas, F., \& Morbidelli, A.\ 1996, Celestial Mechanics and Dynamical Astronomy, 64, 209 
\item Trujillo, C.~A., Jewitt, D.~C., \& Luu, J.~X.\ 2000, \apjl, 529, L103 
\item  Tsiganis, K., Gomes, R., Morbidelli, A., \& Levison, H.~F.\ 2005, \nat, 435, 459 
\item Vokrouhlick{\'y}, D., Pokorn{\'y}, P., \& Nesvorn{\'y}, D.\ 2012, \icarus, 219, 150 
\item Wetherill, G.~W.\ 1967,  \jgr, 72, 2429 

\end{itemize}

\end{document}